# PoultryTalk: A Multi-modal Retrieval-Augmented Generation (RAG) System for Intelligent Poultry Management and Decision Support

Kapalik Khanal[1], Biswash Khatiwada[1], Stephen Afrifa[1], Ranjan Sapkota[2], Sanjay Shah[1], Frank Bai[1], Ramesh Bahadur Bist[1,*]

[1]*Department of Biological and Agricultural Engineering, North Carolina State University, Raleigh, NC, USA, 27606*

[2]*Department of Biological and Environmental Engineering, Cornell University, Ithaca, NY 14850, USA*

**Corresponding: rbbist@ncsu.edu*

**ABSTRACT** The Poultry industry plays a vital role in global food security, yet small- and medium-scale farmers frequently lack timely access to expert-level support for disease diagnosis, nutrition planning, and management decisions. With rising climate stress, unpredictable feed prices, and persistent disease threats, poultry producers often struggle to make quick, informed decisions. Therefore, there is a critical need for intelligent, data-driven systems that can deliver reliable, on-demand consultation. This paper presents PoultryTalk, a novel multi-modal Retrieval-Augmented Generation (RAG) system designed to provide real-time expert guidance through text and image-based interaction. PoultryTalk uses OpenAI's text-embedding-3-small and GPT-4o to provide smart, context-aware poultry management advice from text, images, or questions. System usability and performance were evaluated using 200 expert-verified queries and feedback from 34 participants who submitted 267 queries to the PoultryTalk prototype. The expert-verified benchmark queries confirmed strong technical performance, achieving a semantic similarity of 84.0% and an average response latency of 3.6 seconds. Compared with OpenAI's GPT-4o, PoultryTalk delivered more accurate and reliable information related to poultry. Based on participants' evaluations, PoultryTalk achieved a response accuracy of 89.9%, with about 9.1% of responses rated as incorrect. A post-use survey indicated high user satisfaction: 95.6% of participants reported that the chatbot provided "always correct" and "mostly correct" answers. 82.6% indicated they would recommend the tool, and 17.4% responded "maybe." These results collectively demonstrate that PoultryTalk not only delivers accurate, contextually relevant information but also demonstrates strong user acceptance and scalability potential.

## 1. Introduction

Agriculture is a key foundation of global food security and economic stability, supporting the livelihoods of billions (Robert, 2008). In agriculture, poultry production is one of the fastest-growing and most economically important sectors of livestock farming (Mottet and Tempio, 2017). It helps provide a critical source of affordable, high-quality protein. Poultry production boosts rural incomes and diversifies farm resources, making it a key component of sustainable agriculture (Bist et al., 2024). However, producers face challenges, including flock health, nutrition, climate change, emerging diseases, fluctuating input costs, and market volatility. Poor nutrition and housing harm both productivity and animal welfare (Phillips, 2016). Many farmers lack reliable, timely expertise and rely on informal sources (Šūmane et al., 2018), which deepens the gap between advances in poultry science and real-world use. This knowledge gap widens the disconnect between advances in poultry science and their practical application on farms, limiting productivity, profitability, and sustainability. Therefore, research-based knowledge about poultry is fundamental.

In response, agricultural extension services have served as the primary link between research and practical application (Aflakpui, 2007; Cash, 2001). Extension agents guide farmers through on-site visits, workshops, and informational bulletins. However, the global shortage of trained extension agents has become a significant challenge (Belay and Abebaw, 2004; Saleh et al., 2016). Those who were available often lack round-the-clock accessibility and are difficult for small and medium farmers to reach. In addition, poultry producers were widely scattered across rural and remote regions, which makes this extension service less practical and sustainable (Mapiye et al., 2021). Furthermore, farmers often face language barriers, inconsistent internet access, and lack the time to read lengthy documents (Gupta et al., 2024; Prajapati et al., 2025). This highlights the urgent need for digital tools that deliver validated knowledge in poultry science in accessible, actionable formats at the point of need 24/7.

Recent AI advances have enabled new ways to share agricultural knowledge using Large Language Models (LLMs) like ChatGPT (OpenAI, 2025) and Gemini (Google, 2025), which can understand and generate human-like text (Campesato, 2024). This supports interactive learning and knowledge exchange. Generic LLMs are trained on large, often unverified internet datasets (Alber et al., 2025). Therefore, LLMs have significant limitations in providing the most accurate information for domain-specific questions in the poultry domain. This can make them give plausible but incorrect or misleading information called hallucination. For example, an LLM might recommend incorrect feed formulation or misidentify signs of poultry disease. Such mistakes can harm animal welfare, farm income, and food safety. LLMs also cannot access newly published studies or reliably incorporate new research and local practices (Sallam, 2023). Their advice can quickly become outdated or irrelevant. While these LLMs are valuable for learning, they are risky for expert agricultural decisions. It’s always best to cross-check their guidance with up-to-date sources or trusted local experts.

Over the past few decades, agricultural decision-support systems have undergone significant changes (Akaka et al., 2024). Retrieval-Augmented Generation (RAG) architectures address these problems by combining LLMs' language capabilities with the accuracy of information retrieval systems (Balakrishnan and Purwar, 2024; Lewis et al., 2020). Unlike traditional models that rely solely on information stored in their neural networks, RAG systems access explicit memory, such as a dense vector index of Wikipedia via a neural retriever, allowing them to produce more accurate, factual, and context-aware responses (Lewis et al., 2021). They can use the same retrieved passages for an entire sequence or select different passages per token, enhancing flexibility and specificity. Fine-tuned RAG models outperform both parametric-only models and task-specific retrieve-and-extract architectures, generating language that was more precise, diverse, and grounded in evidence (Lewis et al., 2020). In addition, multi-modal inputs (image and text) take these systems beyond basic text-based AI (Yang et al., 2023; Zhou et al., 2022; Zhou and Long, 2023). It makes them a powerful tool for intelligent, data-driven decision support by enabling these systems to understand and reason about visual information alongside written questions.

In agriculture, the RAG model has been widely explored to support farmer decision-making for crops or livestock domains (Jain et al., 2018; Kandamali et al., 2025; Oyelade et al., 2025). For example, FarmChat provides natural language crop advice for seed selection and pest management (Jain et al., 2018), while CottonBot integrates RAG with agentic LLM tools and farm-specific sensors to deliver dynamic, field-specific cotton farming recommendations (Kandamali et al., 2025). Similarly, in livestock, deep learning combined with RAG has been applied for disease detection in cattle, goats, and sheep, providing actionable treatment and prevention advice (Oyelade et al., 2025). Other work has explored LLM-based systems for goat health management, using structured knowledge fusion with table and decision-tree textualization to improve retrieval accuracy (Oyelade et al., 2025). Brazilian veterinary applications integrated RAG with regulatory knowledge bases for context-aware legal guidance (Montero et al., 2025). Multi-modal video-based systems like MooBot demonstrate the utility of converting barn video data into structured insights for precision livestock farming (Raskar et al., 2025). LLMs have been extended to interact with numerical dairy datasets using RAG and NL2SQL approaches. Despite major progress in AI and RAG-based decision-support systems for agriculture and livestock, poultry farming still faces two key gaps. First, existing RAG approaches are not designed around the unique complexities of poultry housing and management. Second, most AI tools support only text-based interaction, limiting their usefulness for farmers who rely on visual inputs or have varying levels of digital or language literacy.

Building on this foundation, this study presents PoultryTalk, a novel multimodal RAG system designed to deliver reliable, evidence-based decision support for the poultry industry. Its development was guided by practical needs, drawing from producer challenges, extension priorities, and observations from real on-farm conditions. The key objectives of this study can be summarized in the following specific points:

1. Design and implement a multi-modal RAG architecture that integrates text and image modalities for comprehensive poultry consultation.
2. Construction of a robust, domain-specific knowledge base that ensures factual, evidence-based responses grounded in authoritative poultry science.
3. Evaluation of the system by measuring retrieval precision, semantic similarity, and response accuracy, with real-world user validation involving farmers, researchers, and industry professionals.
4. Benchmarking and evaluation of the RAG system's outputs against ChatGPT to determine improvements in relevance and factual accuracy.

## 2. Methodology

### *2.1. PoultryTalk RAG System Design*

This section includes the comprehensive methodological framework adopted for the development and evaluation of PoultryTalk. It is a multi-modal RAG system designed to provide intelligent, evidence-based consultation, information, and decision-making in all aspects of poultry farming. The system integrates modules for housing, nutrition, breeding, genetics, welfare, health management, behavior, and precision poultry production. The methodology consists of six major components: system architecture overview, knowledge base construction and validation, multi-modal input processing, retrieval engine design, contextual response generation, and experimental evaluation protocols. Each component was meticulously developed to ensure the system remains accurate, responsive, and practical for poultry stakeholders, including farmers, researchers, and industry experts.

#### *2.1.1. System Architecture Overview*

The overall architecture of PoultryTalk was designed as a modular, multi-layered system that can be visualized as a sequential flowchart (**Fig. 1**). The proposed framework makes it easier to process a wide range of user interactions. It handles both textual and visual inputs, which helps support effective and adaptive communication in practical poultry consultation environments. The architecture begins with an input acquisition layer that captures user queries and converts them into a standardized format suitable for processing. Inputs first go to the multi-modal preprocessing module. Computer vision models analyze image data to extract features. Text inputs were broken down and embedded for context. The data then enters the RAG-based reasoning core, which retrieves information from a poultry knowledge base. Finally, an AI model produces responses that are relevant and supported by evidence. This ensures that each user interaction is both scientifically grounded and tailored to the specific management or diagnostic need. The system also includes a latency optimization and feedback layer, designed to maintain computational efficiency even under variable connectivity conditions common in rural areas. Finally, the output layer delivers the response in a user-friendly format, supporting text and image outputs for inclusivity. The system helps solve important problems in agricultural AI, such as handling different situations, working with various types of data, and adjusting quickly.

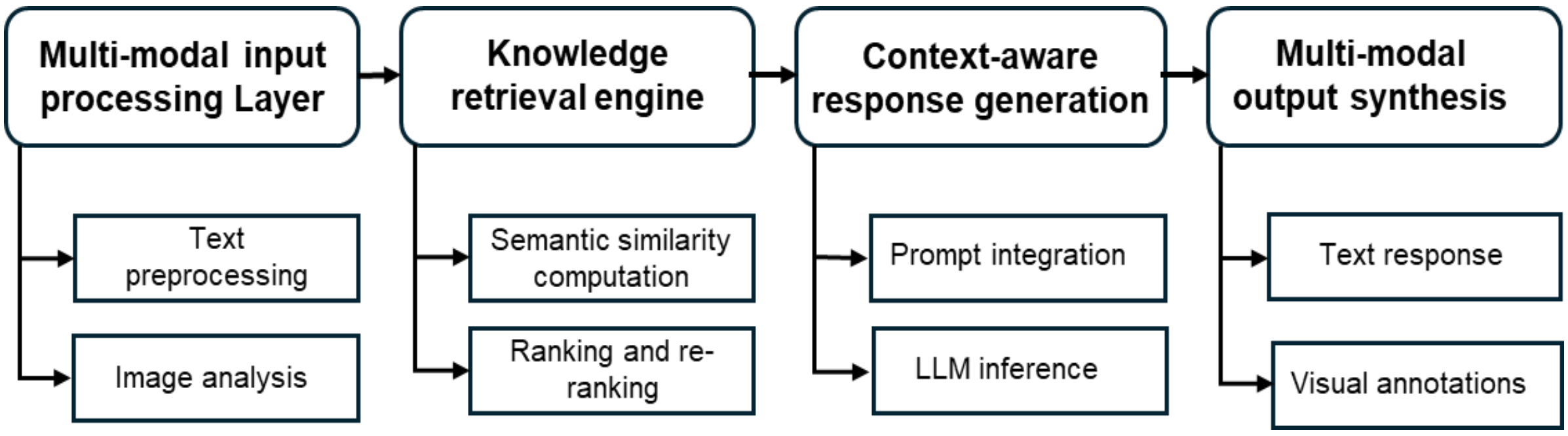

**Fig. 1.** Integrated system architecture and information flow of the PoultryTalk decision-support platform.

*2.1.2. Knowledge Base Construction and Validation*

*2.1.2.1. Document Collection*

The PoultryTalk knowledge base was developed by compiling high-quality content from reputable sources in poultry science. The collection comprised over 462 documents, totaling more than 2,500 pages. The majority of the open source articles database originates from universities such as NC State University, the University of Georgia, the University of Arkansas, Mississippi State University, Auburn University, and Purdue University, including their extension publications. Additionally, 155 open-source, peer-reviewed research articles were incorporated from established journals such as Poultry Science, Computers and Electronics in Agriculture, Animal, and AgriEngineering. Each document was evaluated and sorted out by the expert based on scientific rigor, publication types, and practical relevance to on-farm poultry challenges. Drawing from diverse sources, the resulting knowledge base was both balanced and reliable, enabling the AI system to deliver accurate and relevant responses across various poultry production systems. Furthermore, 200 ground-truth data points were collected from experts to test PoultryTalk and compare its performance with GPT-4o.

*2.1.2.2. Document Processing Pipeline*

To retrieve information efficiently and accurately, this study processed all collected documents through a multi-stage pipeline. First, we used the Py-PDF-Directory-Loader module from LongChain to extract text from PDFs (LongChain, 2025). This helped preserve the necessary structural metadata, which maintains the logical flow and supports precise referencing in later steps. Next, the Recursive-Character-Text-Splitter was used to break large documents into smaller, consistent sections (Pipitone and Alami, 2024). Each chunk was set to 800 characters with an 80-character (10%) overlap to keep information connected between segments. The splitting method was designed to follow natural language patterns, focusing on paragraph breaks, sentence structure, and domain-specific technical terms. This process created a well-organized, searchable dataset ready for vectorization and embedding. As a result, the PoultryTalk system can quickly deliver clear, relevant answers to user questions.

*2.1.2.3. Embedding Generation and Storage*

Once the documents were processed, each text segment was turned into a high-dimensional vector using the OpenAI text-embedding-3-small model. This model creates 1,536-dimensional vectors that capture the meaning and context of the text (Nguyen et al., 2025). The vectors were then stored in ChromaDB, a database designed for vector embeddings. Each segment was labeled with details like its source, publication date, topics, and relevance score.

This extra information will help users search more effectively and identify trusted sources. It also makes it easier to track and check information. With this setup, users can find accurate and relevant guidance based on reliable research and extension materials.

#### *2.1.3. Multi-modal Input Processing*

The proposed multi-modal input processing framework combines text and images to improve poultry health assessment and management. Text queries were prepared by breaking words into parts using farm-related terms, identifying keywords such as species, diseases, and management topics, and sorting questions into categories such as diagnosis, nutrition, reproduction, or management. Images are analyzed using an image analysis process with GPT-4 Vision (OpenAI, 2025), and the resulting response was later verified in context. This step includes basic image prep, automatic identification of disease signs, body condition, and environmental details. Results from both text and images were combined to provide insights grounded in the full context.

#### *2.1.4. PoultryTalk Retrieval Engine Design*

The retrieval engine in PoultryTalk uses a hybrid retrieval mechanism that combines semantic similarity with traditional keyword-based relevance to improve query-document matching. Each user query was compared to documents using a weighted combination of embedding-based similarity and a term-frequency based score (BM25) in equation *(i)* (Rayo et al., 2025).

$$Score\ (q,d) = \alpha \times SemanticSim(q,d) + (1-\alpha) \times LexicalSim(q,d) \qquad (i)$$

Here, SemanticSim(q,d) represents the cosine similarity between query and document embeddings, LexicalSim(q,d) was the BM25 relevance score, and $\alpha \in [0,1]$ controls the balance between semantic and lexical information. $\alpha = 0.70$ to emphasize semantic matching while keeping lexical search relevant. This approach ensures that the system captures both the meaning of the query and the exact keyword matches, improving overall retrieval quality.

After computing initial scores, the top-k documents (k = 6) are selected and further re-ranked to enhance diversity while maintaining relevance. The re-ranking step was inspired by the Maximal Marginal Relevance (MMR) framework (Carbonell and Goldstein, 1998), which reduces redundancy in the results. Each document was scored using Goldstein and Carbonell (1998) in equation *(ii)*.

$$FinalScore\ (d) = Score\ (d) - \lambda \times max\ (Sim\ (d,di))\ \forall\ di \in S \qquad (ii)$$

where S was the set of already selected documents and Sim(d,di) was the cosine similarity between document embeddings. This ensures that the retrieved set of documents was both relevant and diverse, covering a broader range of topics to provide actionable and informative responses for farmers.

#### *2.1.5. Response Generation Framework*

The Response Generation Framework was the main engine behind PoultryTalk. It takes user input and generates clear, evidence-based answers for poultry-related queries (**Fig. 2**). The system combines advanced language models with a specialized retrieval process, ensuring responses are both accurate and easy to understand (Huang and Huang, 2024). Unlike typical chatbots that rely only on pre-trained data, PoultryTalk uses a carefully selected set of knowledge-based poultry science documents. For example, it analyzes text for intent and key terms and uses a vision-based classifier to interpret images, identifying features such as lesions, feather loss, or unusual environmental conditions. All inputs were converted into structured text, which the system then uses to find information and generate responses.

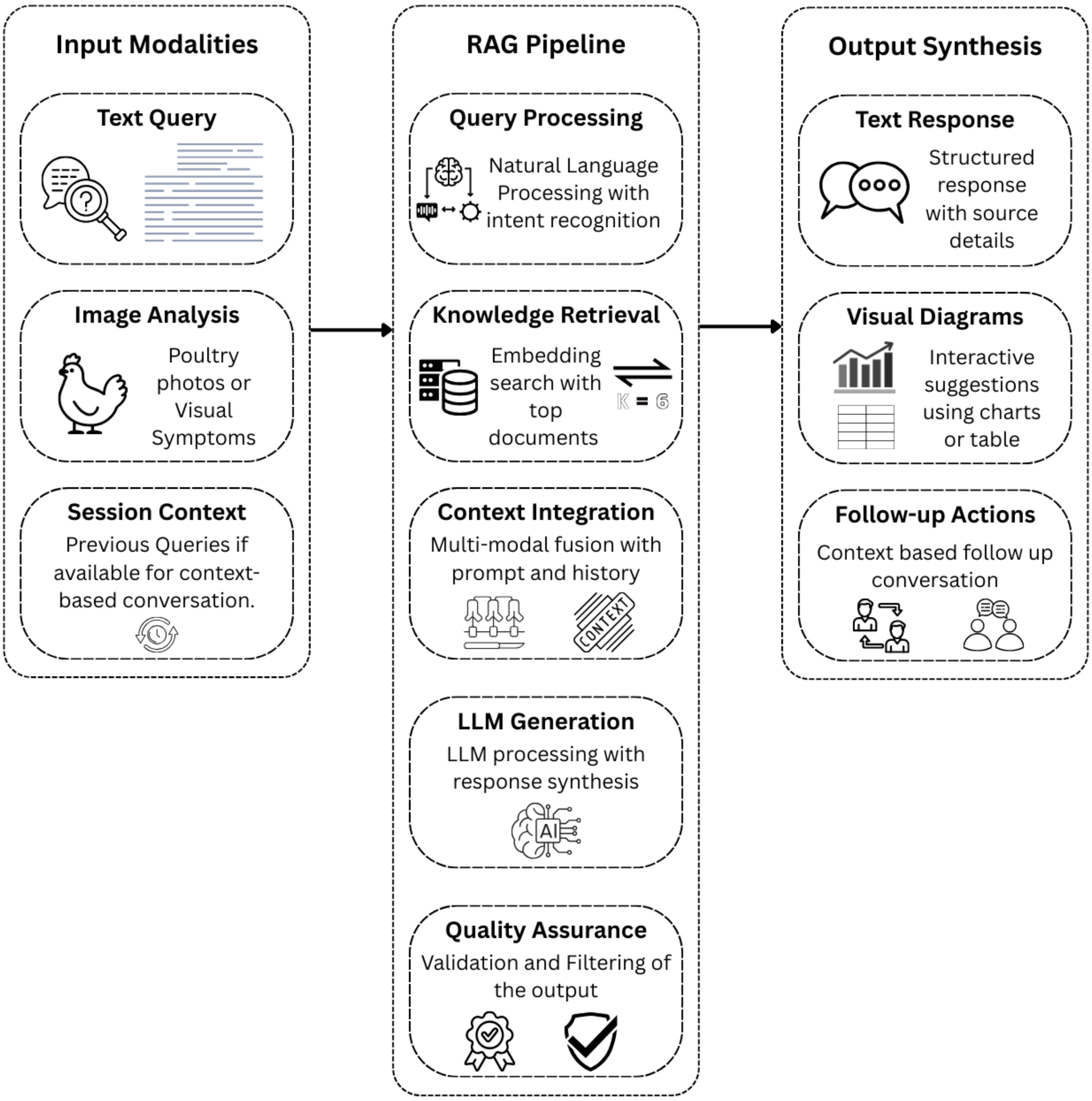

**Fig. 2.** System architecture of the PoultryTalk multimodal response generation framework.

*2.1.5.1. Input Integration*

The first step in the PoultryTalk response pipeline was Input Integration. Here, different types of user inputs were converted into formats the system can understand. For text queries, a specialized module breaks down the language and identifies the user's intent, making it easier to recognize technical poultry terms, management situations, and disease-related keywords. This careful processing helps the system interpret even complex or incomplete questions, reducing errors and improving its ability to find the correct information. When users submit images, GPT-4's vision look for visible symptoms, environmental details, and unusual behaviors in poultry. The system notes features like feather quality, lesions, posture issues, and signs of illness, then turns these observations into descriptive text. Both text and image information were organized in a structured format that maintains their meaning and supports accurate information retrieval. The user's question was combined with previous conversation history using a set template, which helps maintain context and ensures smooth, ongoing interactions (Nguyen et al., 2025). This way, PoultryTalk can give answers that reflect both earlier discussions and new information. The template also guides the system to provide clear, accurate responses based on verified details, helping avoid confusion and repetition. User questions and related documents were added to the conversation history in a structured as shown below;

> *Respond to the question based on the provided context and conversation history. Be concise and accurate. If the question refers to previous messages, use the history to provide context.*
>
> *Conversation History:*
> *{history}*
>
> *Relevant Contexts from Knowledge Base:*
> *{contexts}*
>
> *Question:*
> *{user's question}*

*2.1.5.2. Context Fusion*

Once the inputs were standardized, the next critical step was to combine the user's query and prior conversation history with relevant knowledge from the database. The system retrieves the top six documents from the ChromaDB vector database using OpenAI's text-embedding-3-small (Nguyen et al., 2025). These retrieved segments were then integrated with the structured input using a hybrid scoring mechanism that balances relevance and diversity to optimize contextual responses. The retrieved segments were then integrated with the structured input using a hybrid scoring mechanism that evaluates both relevance and diversity. Relevance ensures that the most applicable scientific and practical information is prioritized, while diversity prevents redundancy by introducing complementary knowledge segments. This approach ensures that the responses reflect a broad yet targeted perspective, encompassing multiple aspects of poultry management, including nutrition, disease prevention, welfare, and housing. Context fusion also leverages conversation history, allowing the system to handle follow-up questions, clarifications, or iterative problem-solving (Watson Benjamin et al., 2025). For example, if a user previously asked about "greenish droppings in layers" and later follows up with "could this be due to diet?", the system automatically references the prior discussion to provide a coherent and progressive answer. By combining retrieved knowledge, user input, and historical context, the framework creates a rich, structured information environment for the generative model to operate effectively.

*2.1.5.3. Response Synthesis*

The Response Synthesis stage was where the integrated information was transformed into a human-readable answer (Mao et al., 2025). GPTs generate responses using a structured prompt template that incorporates user intent, fused knowledge-retrieval context, and prior conversation history. The template explicitly instructs the model to prioritize conciseness, accuracy, and domain-specific terminology, ensuring that the output aligns with both user expectations and scientific validity. When generating responses, GPT aims to do three things: answer questions accurately, offer helpful advice, and use clear, professional language for everyone from farmers to researchers. Its ability to generate modular, adaptable responses enables the system to provide concise summaries, including references or supporting details from the knowledge base when requested. To enhance precision, responses undergo an internal validation step in which the system cross-references key entities, numerical data, and terminology against the retrieved documents. This reduces the likelihood of hallucinations or misinterpretation and ensures that all outputs are firmly grounded in verified agricultural science.

*2.1.5.4. Output Delivery*

The final stage focuses on presenting the generated responses to the user in an accessible and actionable format. Responses were delivered via text on the application interface, ensuring that users can access the information in the mode most convenient to their environment. Lowering the response time allows farmers to receive timely guidance while working in the field. Additionally, the delivery framework supports follow-up questions, iterative problem-solving, and multi-turn conversations, allowing users to engage in natural, human-like dialogue with the system. By combining accurate input interpretation, context-rich fusion, expert-grounded synthesis, and responsive delivery, PoultryTalk provides a complete, end-to-end decision support solution for poultry management. Its designed to delivers scientifically valid, relevant, and actionable interactions that connect advanced AI with on-farm use.

*2.2. Participant Recruitment, System Testing, and User Feedback*

To evaluate the practical utility of PoultryTalk, a diverse group of participants was selected. This group comprised poultry farmers, students, professors, extension agents, and representatives from poultry-related companies. Over 100 individuals (poultry farmers, postgraduate and graduate students, professors, and industry representatives from the USA) were approached to engage with the system, providing a broad spectrum of experience and expertise in poultry management. Participants used PoultryTalk by sending questions through text and images. They asked about topics like housing, nutrition, disease diagnosis, welfare, and production management. During the testing phase, interactions were recorded, and the system's responses were monitored for accuracy, contextual relevance, and responsiveness. After using the system, participants filled out a structured survey in a Google Sheet (**Table 1**). The survey collected both qualitative and quantitative feedback on response quality and overall satisfaction. The survey also gauged participants' likelihood of recommending PoultryTalk to others and their impressions of the multi-modal interface. This approach allowed us to thoroughly assess both how well the system worked and how users responded to it. By testing real-world conditions, we gathered valuable feedback to guide future improvements.

**Table 1.** Survey items and response formats for the PoultryTalk AI ChatBot evaluation.

| Items | Question/Construct | Response Type and Scale |
|---|---|---|
| ***I. Demographic/Background*** | | |
| 1 | Email Address | Required Text Input (Not anonymous) |
| 2 | Background / Role | Multiple Choice (Select one): Master's Student, PhD student, Postdoc, Professor, Industry, Farmer, Other (Please specify) |
| 3 | Name of your University/ Company | Required Short Answer Text |
| ***II. Chatbot Evaluation*** | | |
| 4 | How would you rate this Chatbot overall? | 5-point Likert Scale (1 - Poor to 5 - Excellent) |
| 5 | How would you rate your overall experience with the Chatbot? | 5-point Likert Scale (1 - Poor to 5 - Excellent) |
| 6 | How easy was it to interact with this chatbot? | Multiple Choice Scale: Very Easy, Easy, Neutral, Difficult |
| 7 | Did the chatbot provide correct and relevant answers to your poultry-related questions? | Multiple Choice Scale: Always Correct, Mostly Correct, Sometimes Correct, Rarely Correct |
| 8 | How useful do you find the response for practical poultry management (precision, nutrition, housing, environment, production, health, and disease)? | Multiple Choice Scale: Very Useful, Useful, Somewhat Useful, Not Useful |
| 9 | Which area of the chatbot needs the most improvement? (choose the best one) | Multiple Choice (Select one): Accuracy of Answers, Depth of Knowledge, Clarity of Explanations, Coverage of More Poultry Topics, Speed of Response, Other (Please specify) |
| 10 | What key content of features are missing that should be added to make the Chatbot more helpful? | Short Answer Open Text |
| 11 | Would you recommend this Chatbot to other poultry farmers, students, or professionals? | Multiple Choice (Select one): Yes, No, Maybe |
| 12 | Any additional comments or suggestions? | Optional Open Text |

### *2.3. Experimental Evaluation and Statistical Analysis*

The experimental evaluation of PoultryTalk was designed to assess both its technical performance and user acceptability under realistic usage conditions. The methodology included a combination of controlled model testing, benchmark query assessments, and user-based surveys. During testing, this study asked a series of expert-approved questions on poultry health, nutrition, management, and welfare. This helped show how quickly and accurately the model found information and how well its answers matched what we asked. At the same time, we evaluated how the system worked in real-world settings by giving a structured survey to a wide range of participants. This group included poultry farmers, postgraduate students, professors, and industry stakeholders, all of whom used the system with both text and image inputs. The survey asked participants about how accurate, clear, and relevant they found the responses. It also measured their overall satisfaction and whether they would recommend the system. For statistical analysis, both descriptive and inferential measures were applied. Means and confidence intervals were computed for continuous metrics such as retrieval precision and latency. Categorical survey responses were analyzed using frequencies and percentages to detect significant differences in user perceptions across experience levels and backgrounds. Furthermore, the performances of PoultryTalk and GPT-4o were evaluated against the ground truth.

## 3. Results and discussions

### *3.1.Aggregate Performance Metrics*

**Table 2** shares the results of testing PoultryTalk's ability to find information, answer questions, and respond quickly. PoultryTalk reached an average semantic similarity of 84%, showing it performed well on most questions but less so on others. The system gives clear, accurate answers to simple questions, but it struggles with more complex or multi-step ones. Its retrieval precision was 34%, which means it can find relevant information, though not always with high accuracy. This could be because the model does not always recognize poultry-specific terms or has a limited knowledge base. Improving the model, adding more content, or upgrading its core components could help it answer technical or uncommon questions more effectively (Leite et al., 2025). The average response time was 3.60 seconds. Using methods such as query caching, parallel processing, and local storage can help users get answers faster (Jin et al., 2025) and improve overall performance.

**Table 2.** Summary of aggregate PoultryTalk performance metrics

| Metric | Value | Description |
|---|---|---|
| Number of Questions | 200 | Total ground-truth pairs tested. |
| Average Semantic Similarity | 0.84 | Mean cosine similarity between generated responses and ground-truth answers (higher was better; range 0-1). Indicates response relevance and accuracy. |
| Average Retrieval Precision | 0.34 | Mean similarity score of top retrieved documents to queries (higher was better; range 0-1). Reflects knowledge base effectiveness. |
| Average Latency (seconds) | 3.60 | Mean query processing time, including retrieval and generation. Measures system responsiveness for real-time use. |

### *3.2.Per-Query Insights*

#### *3.2.1. High-Performing Queries*

**Table 3** shows several high-performing queries when comparing the expected and actual responses. For instance, the query "Describe oxidative stress in poultry production? Is it a major concern?" achieved a semantic similarity score of 88%, indicating a high degree of alignment between the generated response and the expert ground-truth answer. In addition, the query "What are the different noises a turkey makes?" achieved a semantic similarity of 92%, demonstrating that the system accurately captured key vocalizations (cluck, putt, cackle, purr, gobble) and provided additional context on behavior and welfare. Similarly, queries on broiler water consumption in poultry production scored above 90%, reflecting the system's strength in well-documented topics in poultry science. This suggests that PoultryTalk excels in retrieving and generating accurate responses for queries grounded in well-documented poultry science domains.

**Table 3.** Some examples of the high-performing queries and corresponding semantic similarity scores in PoultryTalk.

| Test Case | Question | Expected Response | Actual Response (Excerpt) | Correct | Semantic Similarity |
|---|---|---|---|---|---|
| 1 | What are the different noises a turkey makes? | Turkeys make various vocalizations, each serving a specific purpose: 1. Cluck: A soft sound indicating presence or excitement. Practical for farmers to recognize content or alert states. 2. Putt: A sharp cluck signaling alarm. Helpful in identifying stress or threats in the flock. 3. Cackle: Loud, rapid clucks indicating excitement. 4. Purr: A soft sound showing contentment. Indicates a calm environment.<br>4. Gobble: loud, rapid gurgling sound primarily used by tom turkey to attract hens and assert dominance | Turkeys produce a variety of sounds, including gobbles, purrs, yelps, clucks, and putts. Gobbling is most common in males during mating season to attract females and establish dominance. Purrs indicate contentment, while yelps are used for communication within the flock. Clucks can signal reassurance or mild alarm, and putts often indicate danger or alertness. For poultry farmers, understanding these sounds can aid in monitoring turkey behavior and welfare. Recognizing distress calls, such as putts, can help identify potential threats or stressors in the environment. Encouraging a calm environment can reduce stress-related noises, improving overall flock health. Scientific research highlights the importance of acoustic communication in social interactions and stress indicators. Regularly observing and interpreting turkey vocalizations can enhance welfare assessments and productivity. | Yes | 0.92 |
| 2 | Common infectious diseases in backyard poultry? | Common infectious diseases in backyard poultry include Marek’s, Newcastle, Avian influenza, Fowl Pox, and Salmonella. 1. Marek's Disease: A viral disease causing tumors and paralysis. 2. Newcastle Disease: Highly contagious, affecting the | Common infectious diseases in backyard poultry include Salmonella, coccidiosis, and avian influenza. Salmonella can be present in healthy-looking birds and is easily transmitted to humans. Coccidiosis, caused by protozoa, leads to intestinal issues and poor productivity. Avian influenza can cause respiratory symptoms and sudden death. Regularly clean and disinfect equipment and limit contact with wild | Yes | 0.78 |
| | | respiratory, nervous, and digestive systems. 3. Avian Influenza: A viral infection with varying severity. 4. Fowl Pox: Causes lesions on skin and mucous membranes. 5. Salmonellosis: Caused by Salmonella bacteria, leading to diarrhea and mortality. | birds. For coccidiosis, ensure proper litter management and consider using medicated feed if necessary. Although an avian influenza vaccine is available in the U.S., monitoring for symptoms can help control outbreaks. These practices are supported by research emphasizing hygiene and biosecurity to prevent the spread of disease. | | |
| 3 | How much water do broilers drink? | Broilers typically drink about a quart (0.95L) of water for every pound (0.45kg) of feed consumed. With an average feed conversion ratio of 1.75, this equates to roughly half a gallon of water per pound of bird grown. | Broilers typically consume about a quart of water for every pound of feed consumed. With an average feed conversion ratio of 1.75, this equates to nearly half a gallon of water per pound of bird grown. For practical application, ensure drinker systems are well-maintained to provide consistent water flow, as proper hydration is crucial for growth and heat stress management. Scientific research highlights the importance of drinker line management to optimize water intake and flock performance. | Yes | 0.95 |
| 4 | Describe oxidative stress in poultry production? Is it a major concern? | Oxidative stress in poultry production is a critical concern, arising when there is an imbalance between the production of reactive oxygen species (ROS) and the bird’s natural antioxidant defenses. ROS are highly reactive molecules that can damage cellular | Oxidative stress in poultry production is a significant concern, characterized by an imbalance between reactive oxygen species (ROS) production and the bird's antioxidant defenses. This stress can lead to cellular damage, affecting growth, reproduction, and overall health. It is exacerbated by environmental stressors like high temperatures, often linked to climate change. Practical applications for | Yes | 0.88 |

| | | | |
|---|---|---|---|
| | components such as lipids, proteins, and DNA. In poultry, excessive oxidative stress can impair growth, reduce feed efficiency, compromise immune function, and negatively affect reproduction and egg quality. Environmental factors like heat stress, poor nutrition, high stocking density, and disease challenges can exacerbate ROS production, making oxidative stress a common issue in intensive poultry systems. Managing oxidative stress through proper nutrition, antioxidants supplementation, and good husbandry practices is therefore essential for maintaining flock health and productivity. | managing oxidative stress include: 1. Dietary Antioxidants: Supplementing poultry diets with antioxidants such as vitamin E, selenium, and plant extracts can mitigate oxidative damage. 2. Environmental Management: Implementing cooling systems and proper ventilation reduces heat stress, a major oxidative stress contributor. Scientific research underscores the importance of these strategies, highlighting that oxidative stress can decrease meat quality and increase production costs. Recommendations include adopting a multidisciplinary approach, integrating nutritional, environmental, and management strategies to enhance poultry welfare and productivity. Regular monitoring and adjusting practices based on environmental conditions are crucial for sustainable production. | |

### *3.2.2. Lower-Performing Queries*

PoultryTalk struggled with broader or less specific questions. For example, when asked about animal welfare in poultry production or how much water birds drink relative to their feed intake, the answers only partly matched expert responses, with similarity scores ranging from 0.33 to 0.46 (**Table 4**). Even on topics such as precision poultry farming or lighting effects in broilers, similarity remained moderate, ranging from 0.39 to 0.46. This usually happened because the system gave general answers that did not fully address the questions. These results suggest that PoultryTalk works best in well-documented areas, such as housing and management, health, broiler nutrition, and behavior. Expanding its knowledge base and improving how it classifies questions could help it perform better with these types of queries (Leite et al., 2025).

**Table 4.** Lower-performing queries and corresponding semantic similarity scores.

| Test Case | Question | Expected Response | Actual Response (Excerpt) | Correct | Avg. Similarity Score |
|---|---|---|---|---|---|
| 1 | What are the animal welfare issues related to poultry production? | Poultry farms often face challenges with animal welfare, such as overcrowding, limited movement, and joint problems. Providing good nutrition and making careful breeding choices can help prevent health problems like….. | Broiler production faces several animal welfare issues, primarily due to rapid growth rates. Key concerns include lameness, which restricts movement and causes discomfort, and conditions like footpad dermatitis and hock burns from poor litter management. Overcrowding and limited space can further….. | Partial | 0.41 |
| 2 | How much water does a bird drink compared to the feed they eats? | When it is cooler, birds typically drink about 2 to 3 milliliters of water for each gram of feed they eat. In hot weather, they might drink as much as 4.9 milliliters per gram to stay cool… | Birds typically consume water at a ratio of 2:1 to 3:1 compared to feed intake during cooler temperatures, meaning for every gram of feed, they drink 2-3 ml of water. In hotter conditions, this ratio can increase to 4.9:1, as birds drink more to regulate body temperature (Holik, 2010). Practical applications…… | Partial | 0.33 |
| 3 | What is precision poultry farming? | The application of advanced technologies, including sensors, cameras, and data analytics, to monitor and manage poultry flocks in real time enables….. | Precision Poultry Farming (PPF) is an innovative approach that leverages advanced technologies such as sensors, AI, IoT devices, and robotics to monitor and manage poultry health, welfare, and productivity in real time. 2. Practical... | Partial | 0.39 |
| 4 | What's the impact of using 20 lux light intensity instead of a lower light intensity for my broiler production? | When you set the light in broiler houses to 20 lux, the birds become more active, moving around and exploring more. This helps reduce boredom and supports more natural behaviors, …. | Using 20 lux light intensity for broiler production, compared to lower intensities, can enhance bird welfare and performance. Research indicates that maintaining a minimum of 20 lux improves broiler activity, reduces leg issues, and supports better immune function……. | Partial | 0.46 |

### *3.2.3. Latency Variation*

Query processing latency varied significantly based on query complexity. Simple text-based queries were processed in approximately 2–3 seconds, whereas queries requiring complex context retrieval or potential image processing exhibited latencies up to 11.76 seconds. This variability underscores the computational overhead of multimodal processing and suggests opportunities for optimization, such as streamlining API calls or improving vector database indexing.

### *3.2.4. Context Utilization*

On average, the system retrieved six context documents per query (k=6), with similarity scores ranging from 0.22 to 0.95 (**Fig. 3**). Poultry-specific enhancements, such as keyword boosting, improved document relevance, particularly for queries related to broiler growth and management. For instance, the query on broiler growth rates retrieved contexts with an average similarity of 0.67, citing sources like the Poultry Extension Collaborative (PEC) newsletters. This indicates that the system's retrieval mechanism was effective for specialized poultry topics but could benefit from refined embedding models to boost precision across diverse query types.

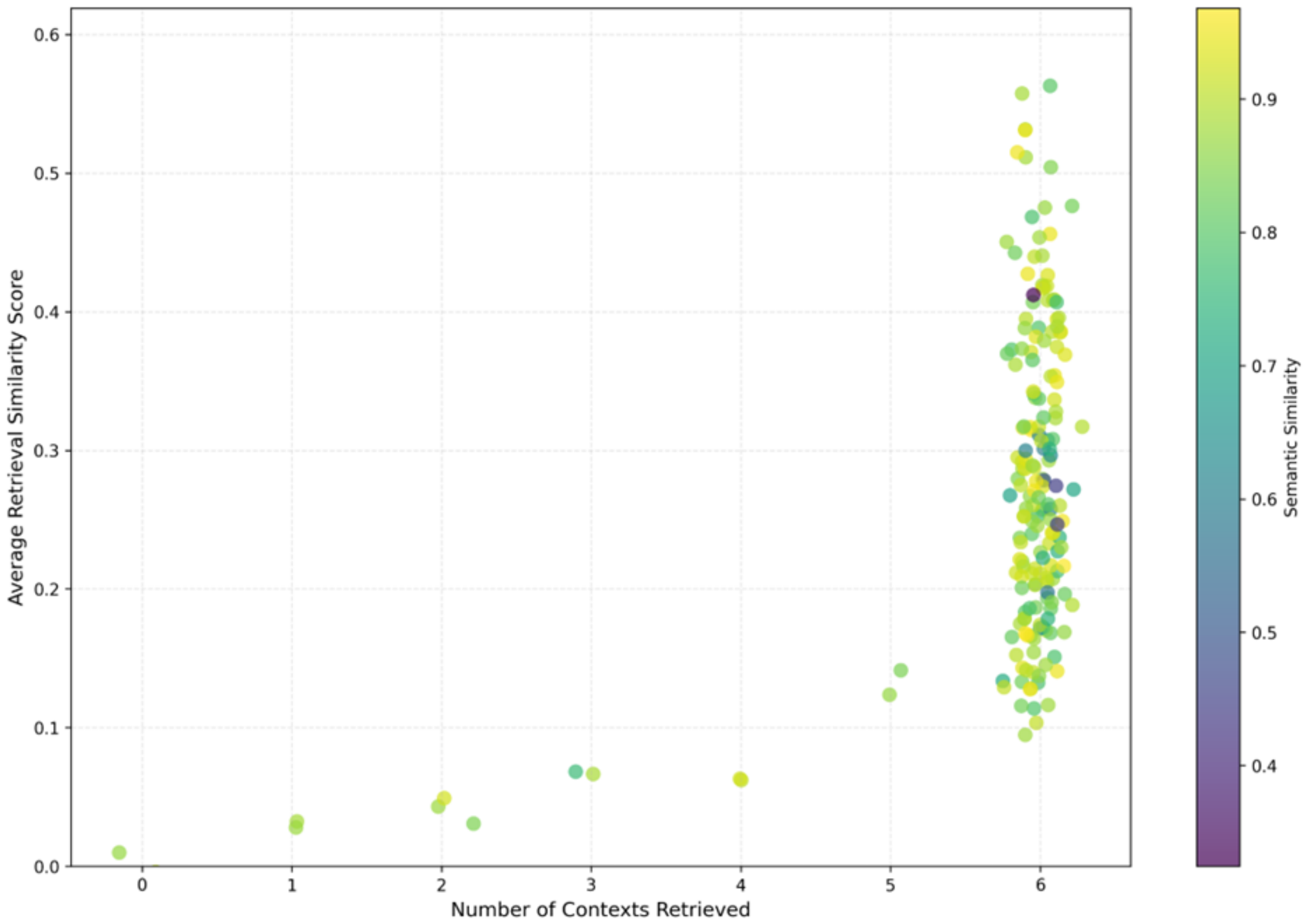

**Fig. 3.** Relationship between number of retrieved contexts and average retrieval similarity score, colored by semantic similarity

*3.2.5. Visualization of Semantic Similarity*

To further characterize response quality, **Fig. 4** illustrates the distribution of semantic similarity scores across all 200 queries. The histogram, overlaid with a fitted normal distribution, indicates that approximately 84% of queries fall within one standard deviation of the mean, suggesting moderate consistency in response accuracy. This visualization underscores the system's balanced performance while highlighting areas for improvement to achieve higher similarity scores across a broader range of queries.

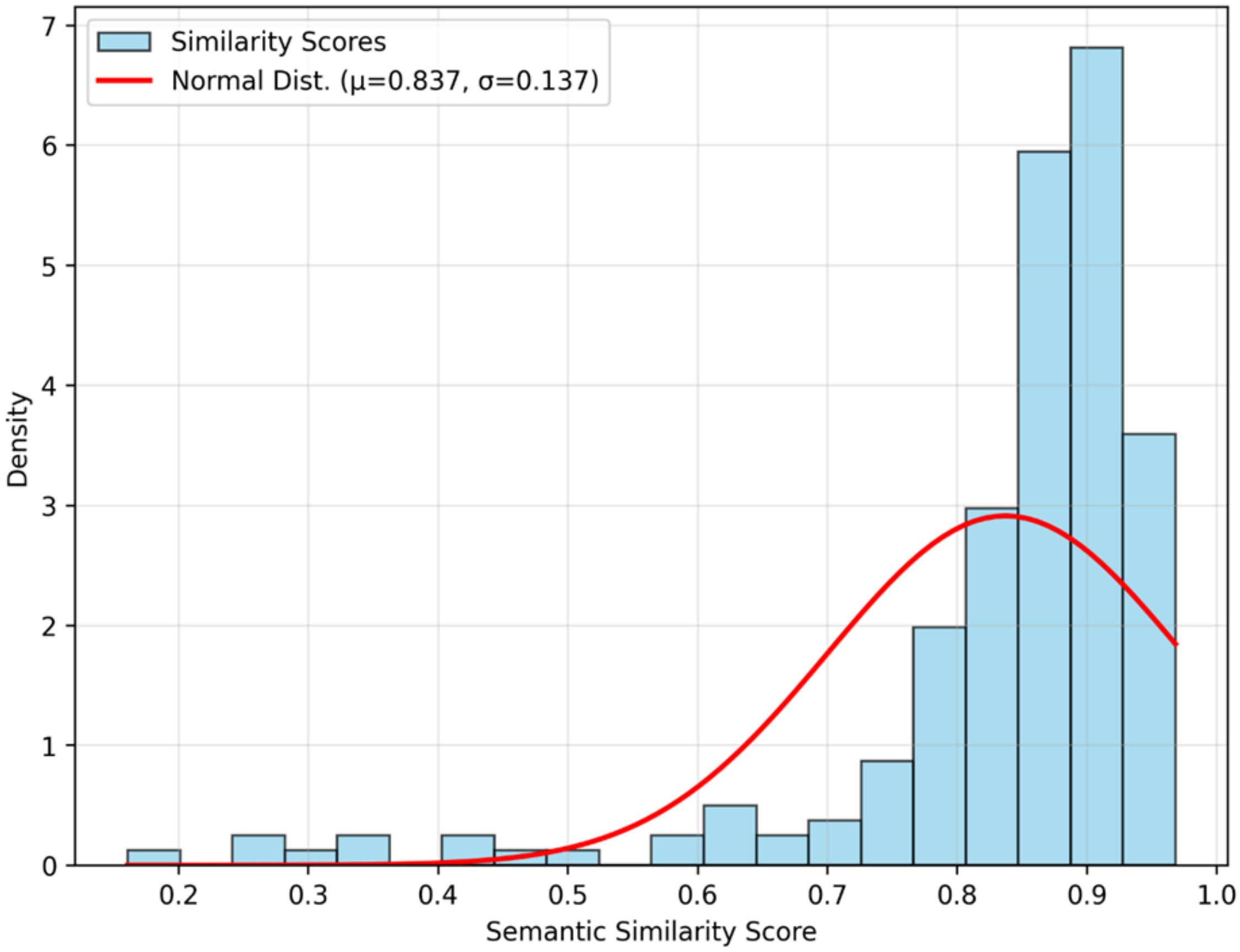

**Fig. 4.** Distribution of semantic similarity scores for PoultryTalk responses.

### *3.3. PoultryTalk Performance Comparison*

To evaluate the relative performance of PoultryTalk compared to a general-purpose LLMs, we used OpenAI GPT-4o as a reference. This model was selected for its advanced natural language understanding and generation abilities, positioning it as a state-of-the-art general AI that functions effectively without domain-specific retrieval support (OpenAI, 2025). The test was conducted using the identical 200 query-answer pairs that experts verified for the ground-truth data set. Evaluation criteria included semantic similarity to expert answers, inter-system answer similarity, and retrieval precision. Each system was given queries independently, and GPT-4o was asked to serve as a knowledgeable assistant on poultry farming to provide a fair comparison.

#### *3.3.1 Aggregate Metrics*

This study presents the overall performance of PoultryTalk and GPT-4o across 200 ground-truth datasets. PoultryTalk and GPT-4o had average semantic similarities of 84% and 81%, respectively, compared to ground-truth responses. PoultryTalk performed higher in semantic similarity than GPT-4o. This demonstrates PoultryTalk's slightly enhanced ability to generate domain-relevant responses to poultry-related queries while reducing hallucinations, thereby enhancing reliability in practical advisory applications (Abo El-Enen et al., 2025; Tonmoy et al., 2024). These results affirm PoultryTalk's competitive accuracy and efficiency gains, aligning with its design goals for scalable, low-latency poultry advisory support.

#### *3.3.2. Response Quality Analysis*

To provide further insights into the quality of the responses, **Table 5** shows representative samples from the dataset, the generated answers from both PoultryTalk and GPT-4o, and the ground-truth references. For example, answering the question "How to make a feed conversion ratio (FCR) of 1 in broiler chicken?", PoultryTalk provided an answer focused on practical nutrition and management measures, with a semantic similarity of 87%, whereas GPT-4o provided an answer with a semantic similarity of 85%. However, the response from PoultryTalk included caveats based on evidence, which increased reliability. Another instance, 'What is the Histomoniasis in Turkeys?', demonstrates that PoultryTalk achieves 95% similarity with the ground truth by using standard wording and a simpler structure. In comparison, GPT-4o was both content-rich and accurate, but its phrasing was slightly longer. This trend favors the PoultryTalk RAG model in terms of reliability for poultry-specific queries, with up to a 10% decrease in potential error associated with knowledge grounding.

**Table 5.** Comparison of sample responses and semantic similarity scores between RAG and GPT-4o.

| **Query** | **Ground Truth (Excerpt)** | **RAG Response (Excerpt)** | **GPT-4o Response (Excerpt)** | **RAG Sem. Sim.** | **GPT-4o Sem. Sim.** | **RAG vs GPT Sim.** |
|---|---|---|---|---|---|---|
| Do you still need to ventilate a small chicken house in cold weather? | Yes, ventilation is essential for small chickens even in cold weather. Proper ventilation helps manage moisture … | Yes, ventilation is necessary even with small chickens in cold weather… | Yes, you still need to ventilate your chicken coop in cold weather, even when raising small or young chickens… | 0.92 | 0.82 | 0.10 |
| How to make a feed conversion ratio (FCR) of 1 in broiler chicken? | Achieving a feed conversion ratio (FCR) of 1 is not practically possible because some energy and nutrients are always lost as heat or through incomplete digestion or used for maintenance. However, you can get as close as possible to this ideal by optimizing …... | Achieving an FCR (Feed Conversion Ratio) of 1 in broiler chickens is highly challenging and typically unrealistic due to biological limits. However, optimizing FCR involves several strategies. Focus on genetics, nutrition, and management. Select high-efficiency breeds like Ross 308, provide…….. | Achieving a feed conversion ratio (FCR) of 1 in broiler chickens is extremely challenging and rarely possible under commercial conditions, but it can be approached through optimized management and nutrition. Key strategies include providing a balanced, high-quality diet rich in energy…… | 0.87 | 0.85 | 0.02 |
| What are the main nutritional requirements of broiler chickens for optimal growth? | Broiler chickens need a balanced diet that includes enough protein, energy from carbohydrates and fats, as well as vitamins and minerals. Protein helps their muscles grow, energy keeps them active, and vitamins and minerals support their health and the development of bones, feathers, and organs. | Broiler chickens require a balanced diet with adequate protein, energy, vitamins, and minerals for optimal growth. Protein supports muscle development, energy fuels growth and activity, and vitamins and minerals maintain health, bone strength, and proper organ function. Proper feed formulation ensures fast…… | Broiler chickens require a balanced diet that provides the essential nutrients for rapid growth and efficient feed conversion. Their main nutritional needs include high-quality protein for muscle development, energy sources such as carbohydrates and fats, and adequate vitamins and minerals for metabolic and immune……… | 0.94 | 0.91 | 0.03 |
| What is the Histomoniasis in Turkeys? | Histomoniasis, also called blackhead disease, is caused by the parasite Histomonas meleagridis. It mainly affects turkeys, leading to liver and gut damage….. | Histomoniasis, also known as "blackhead disease," is a significant disease in turkeys caused by the protozoan Histomonas meleagridis. Turkeys are highly susceptible, often experiencing…. | Histomoniasis (often called Histomonas meleagridis) is a protozoan parasite that causes blackhead disease (histomoniasis) in turkeys. It mainly damages the liver and cecum, leading to weakness, yellow droppings,….. | 0.95 | 0.91 | 0.04 |

### *3.3.3. Content Utilization and Inter-System Alignment*

PoultryTalk retrieved an average of six contextual references per query, supporting answers with poultry expertise. This strong retrieval rate shows that the RAG framework was effective. Both RAG-GPT and PoultryTalk gave similar answers, indicating shared core information, but PoultryTalk's RAG system verifies sources, reducing the risk of unsupported generalizations seen with GPT-4o. While both systems were consistent, PoultryTalk's RAG approach kept answers aligned with trusted poultry knowledge. This makes PoultryTalk reliable for fact-based answers, which were crucial in specialized advisory settings.

### 3.4. *Technical Performance and Model Efficacy*

The evaluation of PoultryTalk's technical performance was conducted under real-world conditions. The focus was on retrieval accuracy, response relevance, multimodal processing, and overall system robustness. A total of 100 participants—including poultry farmers, postgraduate students, professors, and industry representatives—were invited to test the system. Of these, 34 actively engaged by submitting queries via text and images. Over the course of the trial, these users generated 267 unique questions spanning critical areas such as housing, feeding, disease diagnosis, welfare, and biosecurity. Participants were given one month to use PoultryTalk, test its features, and complete the survey. The average context source was 5.9 (**Fig. 5**). Results showed that the system consistently retrieved contextually relevant segments from its poultry-specific knowledge base. Approximately 90% of responses were rated correct by participants, with only 10% classified as incorrect (**Fig. 6**). Most of the incorrect answers were attributed to two main factors: (1) out-of-context questions, where the queries fell outside the system's poultry-specific domain, and (2) temporary processing errors during peak simultaneous usage, which affected retrieval consistency. However, this study also found that participants had corrected their question by adding a domain-specific term (**Fig. 7**). For example, **Fig. 7a** shows a participant asking PoultryTalk about light intensity without specifying the species, which elicited a generic response prompting them to clarify the question for poultry. In **Fig. 7b**, after specifying "broiler," the participant received the correct, targeted answer. This highlights the importance of including the species or type in queries to get accurate responses from the AI. Therefore, these findings underscore the importance of maintaining clear domain boundaries and enhancing server-side optimization to sustain accuracy under high load. Furthermore, text and image input modes performed reliably, but image-based queries occasionally returned less specific answers, particularly for non-classical symptoms or low-quality images. This suggests the need for additional training data.

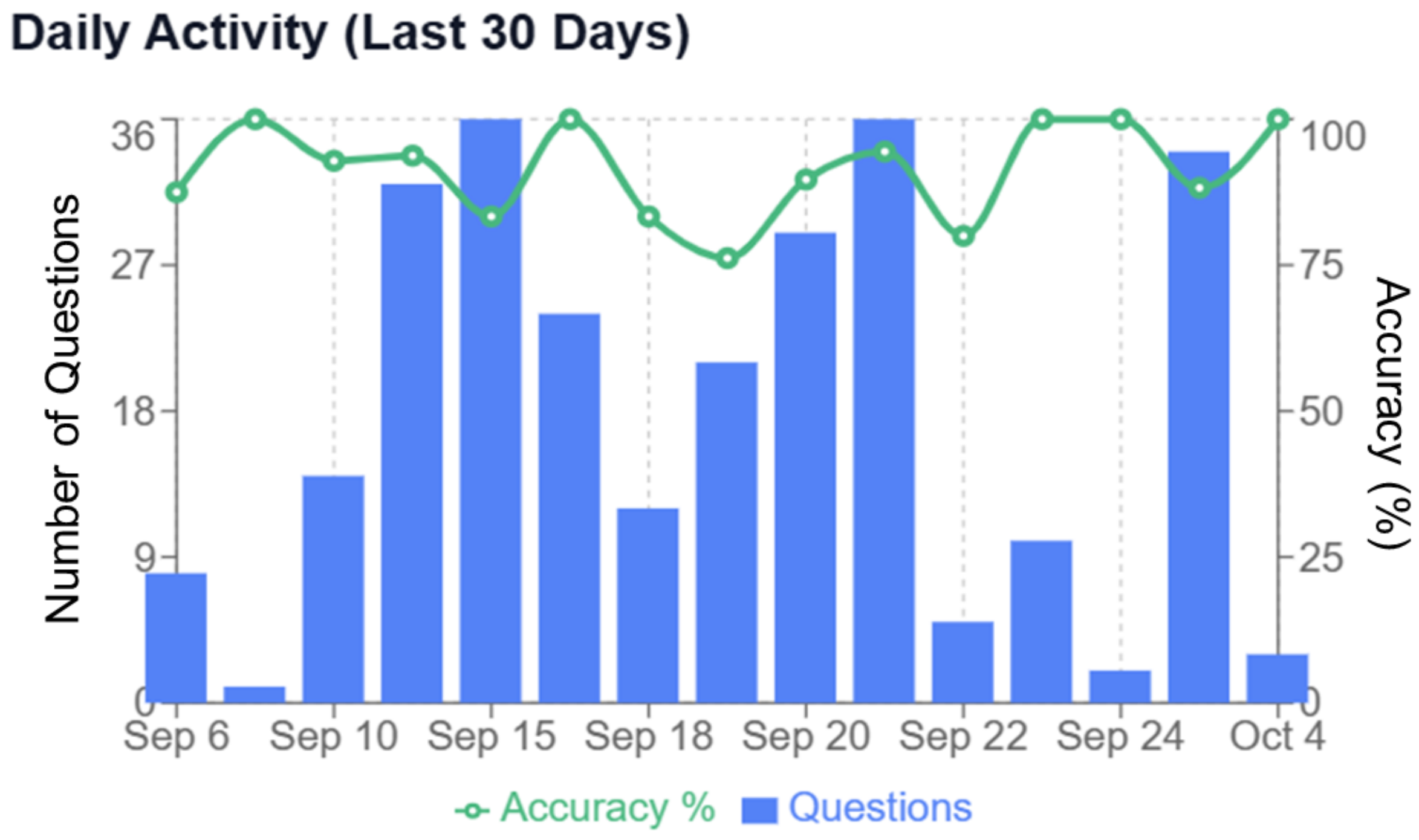

**Fig. 5.** Daily user activity and response accuracy of PoultryTalk over the last 30 Days.

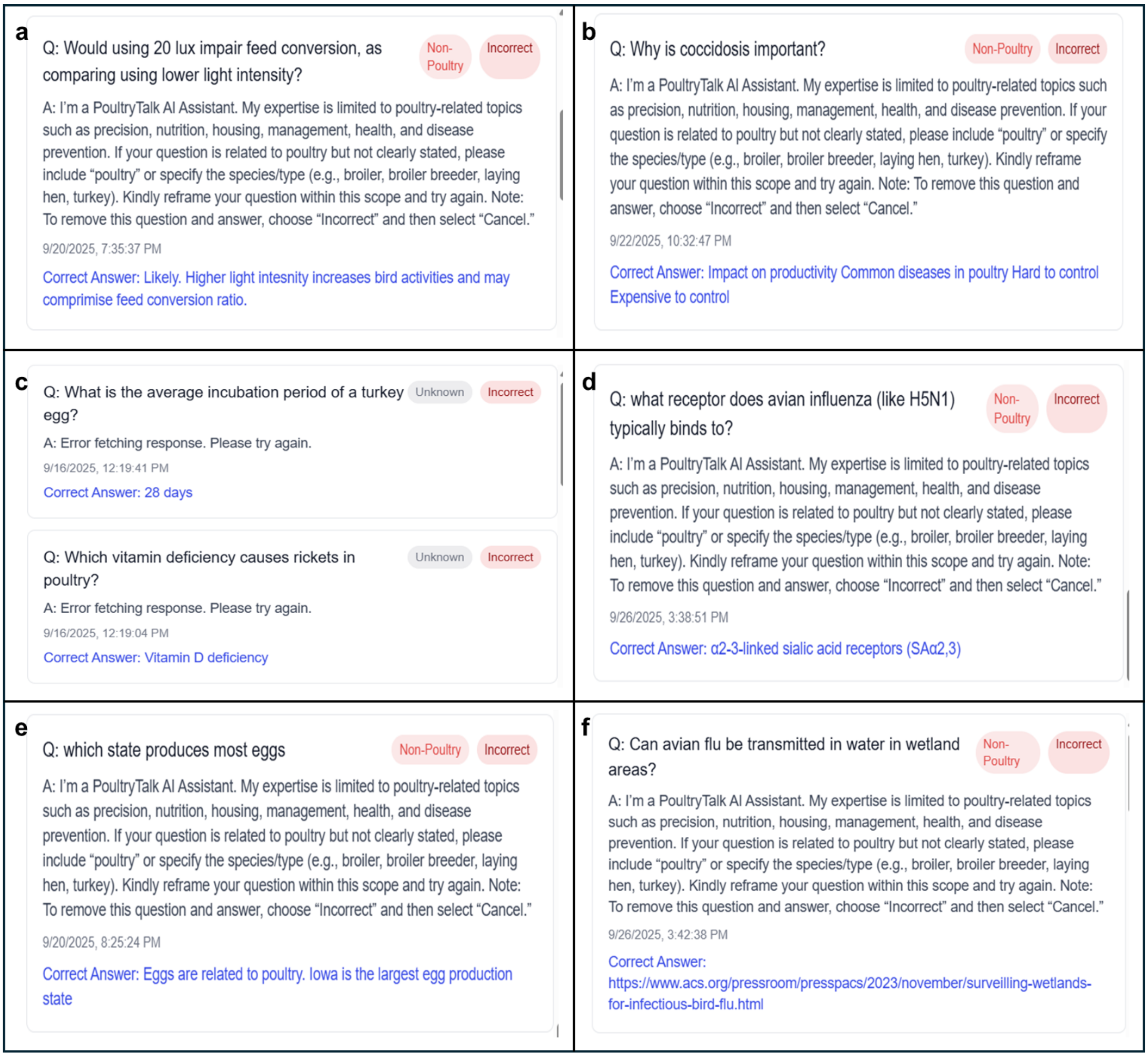

**Fig. 6.** Examples of incorrect or low-accuracy responses in PoultryTalk. The figure illustrates representative cases (a–f) where participants received incorrect or no responses. (a–b) Queries lacked species context (e.g., light intensity or coccidiosis not specified as poultry-related). (c) Network or server overload resulted in a "fetching response" error. (d–f) Ambiguous or missing keywords (e.g., "avian" for influenza or unspecified egg-related queries) caused retrieval mismatches within the RAG framework.

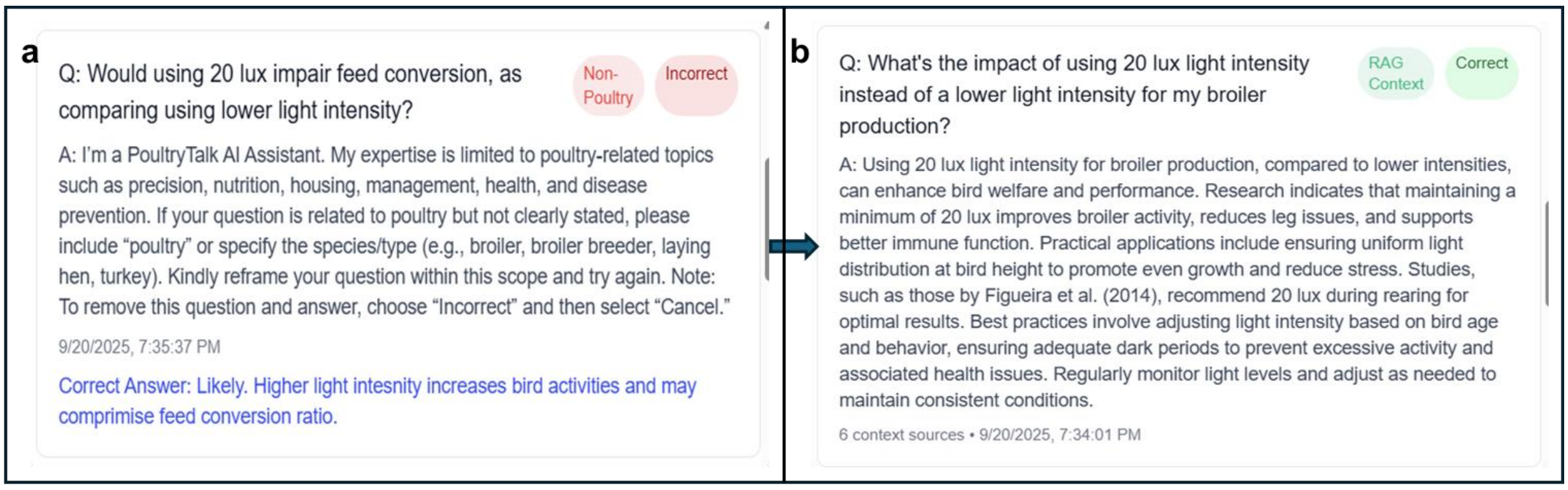

**Fig. 7.** Importance of species specification for accurate PoultryTalk responses. Fig. a) Participant asked about light intensity without specifying poultry type or species; PoultryTalk

returned a clarification prompt. Fig. b) After specifying “broiler,” the participant received the correct answer, showing that clear, species-specific questions improve AI response accuracy.

### *3.5. User Acceptance and Satisfaction (Survey Analysis)*

The user acceptance and satisfaction survey demonstrated that PoultryTalk was perceived as intuitive, effective, and valuable by academic and professional users. Of the over 100 participants invited, 34 interacted with the system and 23 completed the full survey. Respondents included professors, Ph.D. students, postdoctoral researchers, industry professionals, and a master's student, but no poultry producers participated despite being invited. Survey metrics focused on overall rating, user experience, ease of interaction, answer correctness, and practical usefulness (**Fig. 8**). PoultryTalk scored an average overall rating of 3.83 and a user experience score of 3.70, both out of 5, with over 82% rating it as easy to use. More than 95% found their answers to be mostly or always correct. Furthermore, 88% found it helpful for their work or research, 82.6% would recommend it, and 17.4% said "maybe." These strong results underscore PoultryTalk's acceptance and satisfaction among targeted users, reinforcing its potential to support decision-making and knowledge sharing in poultry production. However, the absence of poultry producers in this phase highlights the need for on-farm accessibility, training, localized language options, simplified interfaces, and offline functionality to meet farmers' practical needs and promote broader adoption.

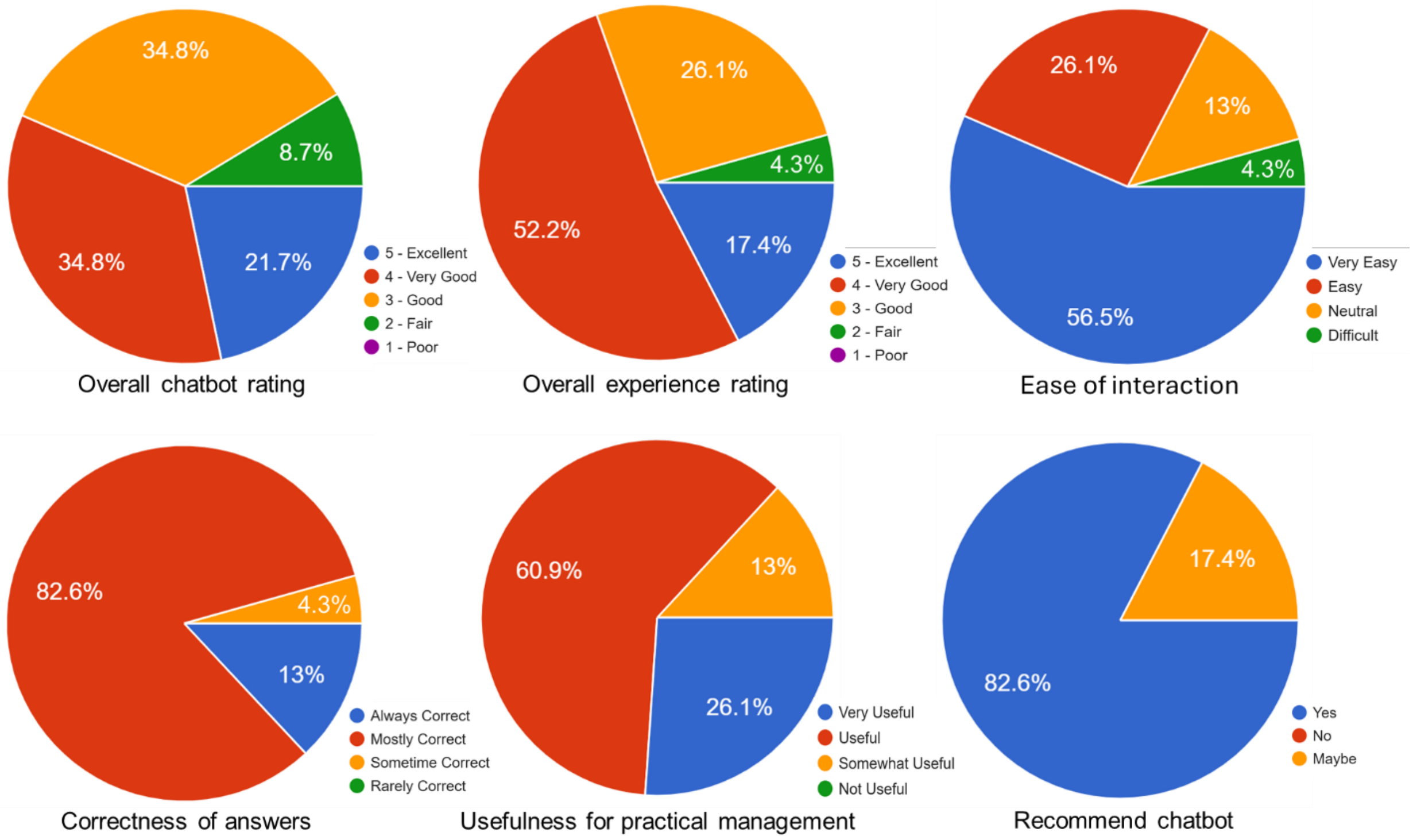

**Fig. 8.** User ratings of PoultryTalk’s overall performance, user experience, usability, accuracy, real-world applicability, and recommendation.

### *3.6. Suggested Improvements and Missing Features (Qualitative Analysis)*

The post-evaluation survey and participants' open-ended feedback provided valuable insights that directly shaped PoultryTalk's iterative improvement. Participants appreciated the system’s concept and its potential to serve as an intelligent, on-demand advisory tool for farmers, educators, and extension specialists. However, several critical areas for enhancement were identified, primarily concerning accuracy, specificity, linguistic understanding, and user experience. Some participants noticed that the system struggled with poultry food safety topics. Users said it often gave incomplete or incorrect answers to questions about food safety, contamination control, or post-harvest management. This issue seems to come from gaps in the current knowledge base. To address this, we plan to add more verified food safety and biosecurity resources to better support all areas of poultry production and handling.

Participants noted the need for automatic spelling correction and better handling of common misspellings and abbreviations. For instance, the system did not recognize typos, such as “nectotic enteritis” instead of “necrotic enteritis,” and sometimes misunderstood abbreviations, such as “HPAI” for “Highly Pathogenic Avian Influenza.” This was because HPAI terms were not in our keyword list and were added later. To solve this, we added an auto-correction and abbreviation-expansion feature to the preprocessing step, so the model can now understand these errors without users having to re-enter their questions. If a participant includes the type of poultry or species in their question, PoultryTalk can also correctly interpret the meaning. Another common concern was that some answers were too general or not specific enough, and sometimes strayed off topic. This happened when the model tried to give more information than the question asked for. To fix this, we improved how responses were generated so answers remain focused, relevant, and specific to poultry. We also upgraded the conversation memory, so PoultryTalk can remember recent exchanges and refer back to them in follow-up questions.

Furthermore, participants requested a more detailed evaluation system (**Fig. 9**), as some answers were only partially correct. In response, we introduced a new scale that measures accuracy in five steps, from 0 to 100 percent, rather than simply marking answers as correct or incorrect. This gives a clearer view of how the model performs. We also learned that users have different preferences for response length. Some prefer short, direct answers, while others want more detailed explanations. To help with this, users can now pick either 'Concise' or 'Detailed' response styles before starting a conversation. We improved citation transparency as well. Many users wanted to see the sources PoultryTalk uses for its answers. Now, the system lists source references or publication titles at the end of each response so that users can check the information. We also noticed that PoultryTalk sometimes tried to answer questions outside the poultry field, which led to mistakes. Since the model was trained only on poultry topics, we added an auto-context flagging feature. If a user asks about something unrelated to poultry, the system now explains its focus and asks the user to rephrase the question to fit within poultry topics. This helps keep answers accurate and guides users to ask relevant questions.

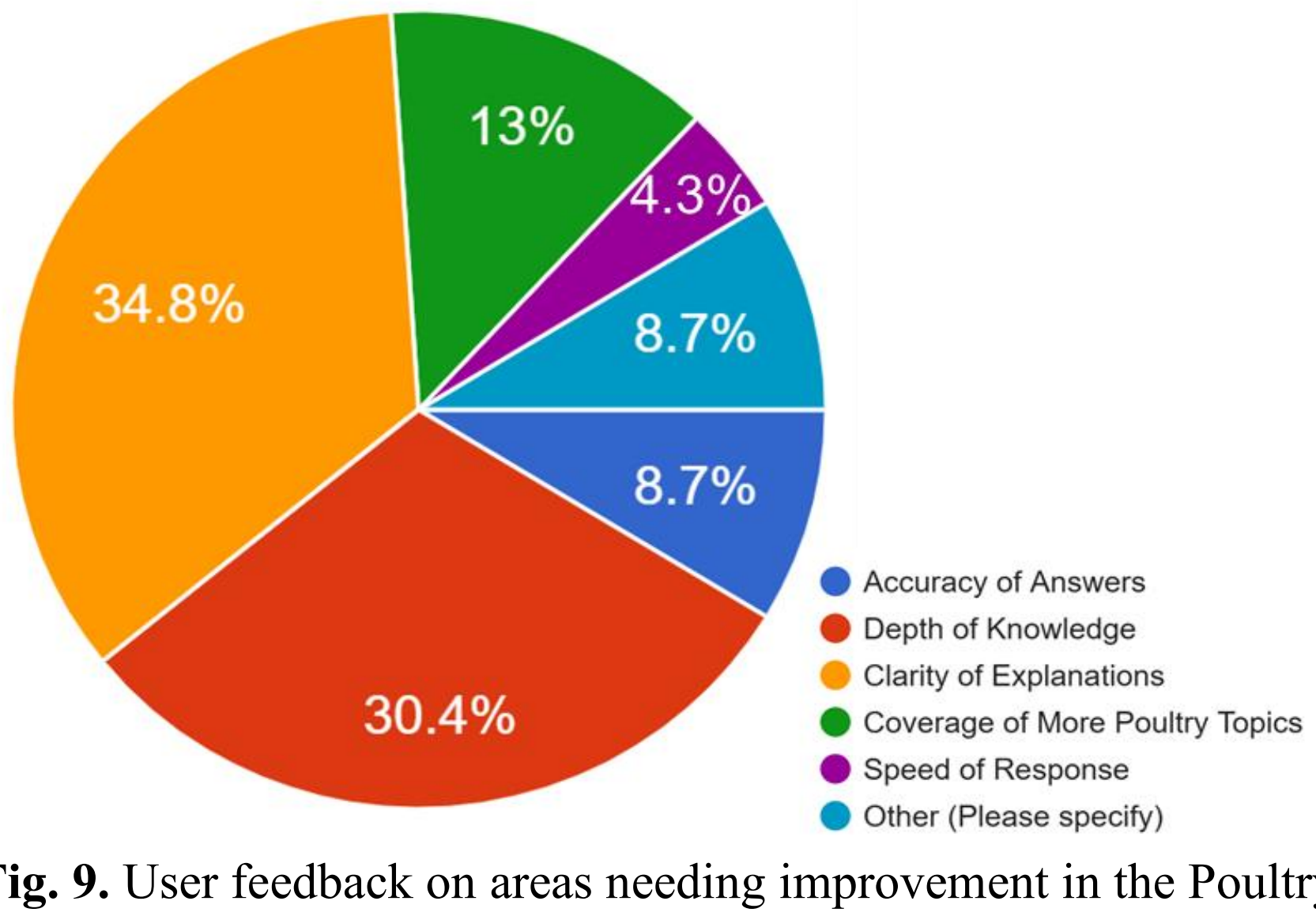

**Fig. 9.** User feedback on areas needing improvement in the PoultryTalk system.

Overall, participants showed strong enthusiasm for PoultryTalk, viewing it as a valuable resource for farmers, students, and poultry professionals. Many praised the project, saying things like “Good going, hats off” and “This could be a very useful tool for poultry producers, educators, and students.” They also suggested improving the user interface, response speed, and visuals, which were now top priorities for the next development phase.

### *3.7. Limitations and Future Directions*

PoultryTalk has shown strong results and was widely accepted by users, but there are still some areas that need improvement. The main issue was that the training data did not fully cover topics like food safety, hatchery management, and new disease monitoring. To address this, future versions will add more information, including food safety standards and recent disease reports, to make the tool more accurate and useful for policy decisions. In addition, the system only retains conversations for a short time and does not learn from users over the long term. To overcome this, subsequent updates should use new learning methods to help PoultryTalk improve based on user feedback and corrections. Furthermore, some users also noticed slow response times and interface problems, especially when the server was busy. Planned improvements, therefore, include redesigning the user interface and making the system faster and more reliable. The team will also work to make PoultryTalk easier to use across devices, especially for mobile users in rural mobile areas with limited internet access. Moreover, multilingual support was another priority, so users can interact in their local languages while still getting accurate technical information. This will help more smallholder farmers who may not speak English. Lastly, future updates also make the tool more transparent and trustworthy by automatically adding references and confidence scores to answers. User feedback has confirmed that PoultryTalk was a valuable tool and has helped shape its future plan. By expanding the data, improving language support, updating the interface, and enabling continuous learning, PoultryTalk aims to become a key resource for the global poultry industry and help connect research with real-world needs.

## 4. Conclusion

PoultryTalk is a RAG-based AI system designed to deliver precise, reliable, and fast decision support for poultry housing and management. It outperforms GPT-4o in poultry topics. It will empower users in resource-limited contexts with expert-validated, poultry-specific guidance. In tests with ground truth data, it achieved 84.0% semantic similarity and a 3.6-second response time. Response accuracy was 89.9%, with 95.6% of users reporting mostly or always correct answers and 82.6% recommending the tool. Errors mainly occurred outside its designed scope or with heavy simultaneous use. Still, PoultryTalk stands out as a user-accepted, targeted solution for precision poultry farming, showing particular strength on poultry-specific queries while being less accurate on broader topics. To further increase reliability and accessibility, future updates should refine retrieval, improve speed, expand the knowledge base, support image queries, and include field trials. These improvements can ensure that PoultryTalk meets the growing demand for data-driven agricultural tools that support sustainable poultry production.

**Acknowledgments**
This project was supported by North Carolina State University’s College of Agriculture and Life Sciences and Department of Biological and Agricultural Engineering Faculty startup fund.

**CRediT authorship contribution statement**
**Kapalik Khanal:** Writing – review & editing, Writing – original draft, Investigation, Visualization, Validation, Supervision, Software, Methodology, Formal analysis, Data curation, Conceptualization. **Biswash Khatiwada:** Writing – review & edit, Investigation. **Stephen Afrifa, Ranjan Sapkota, Sanjay Shah, Fank Bai:** Writing – review & edit, Visualization. **Ramesh Bahadur Bist:** Writing – review & editing, Writing – original draft,

Investigation, Visualization, Validation, Software, Methodology, Formal analysis, Data curation, Conceptualization.

**Declaration of competing interests**
The authors declare that no competing financial interests or personal relationships could have influenced the work presented in this paper.

**Declaration of Artificial intelligence use and integration**
PoultryTalk was developed using GPT-4o as the core language model within a RAG framework to ensure accurate and context-aware responses. In addition, Grammarly was employed to refine and enhance the clarity and consistency of written outputs.